# Liu-type Negative Binomial Regression: A Comparison of Recent Estimators and Applications

Yasin Asar

Department of Mathematics-Computer Sciences, Necmettin Erbakan University, Konya 42090, Turkey, yasar@konya.edu.tr, yasinasar@hotmail.com

**Abstract**

This paper introduces a new biased estimator for the negative binomial regression model that is a generalization of Liu-type estimator proposed for the linear model in [12]. Since the variance of the maximum likelihood estimator (MLE) is inflated when there is multicollinearity between the explanatory variables, a new biased estimator is proposed to solve the problem and decrease the variance of MLE in order to make stable inferences. Moreover, we obtain some theoretical comparisons between the new estimator and some others via matrix mean squared error (MMSE) criterion. Furthermore, a Monte Carlo simulation study is designed to evaluate performances of the estimators in the sense of mean squared error. Finally, a real data application is used to illustrate the benefits of new estimator.

**Mathematics Subjects Classification**: Primary 62J07; Secondary 62J02

# 1. Introduction

In real life contexts, the observations are not independent and identically distributed (iid) all the time. The data often comes in the form of non-negative integers or counts which are not iid. The main interest of a researcher may depend on the covariates which are assumed to affect the parameters of the conditional distribution of events, given the covariates. This is generally achieved by a regression model of count [3]. Thus, count regression models such as Poisson regression or negative binomial (NB) regression are mostly used in the field of health, social, economic and physical sciences such that the non-negative and integer-valued aspect of the outcome plays an important role in the analysis.

Although twenty two different versions of NB model are mentioned in [5], the traditional NB model which was symbolized as NB2 in [2] is the main topic of this paper. This model is more useful than Poisson regression model since NB2 allows for random variation in the Poisson conditional mean, $h_i$, by letting $h_i = z_i \mu_i$ where $\mu_i = \exp(x_i \beta)$ such that $x_i$ is the ith row of the data matrix $X$ of order $n \times (p+1)$ with $p$ explanatory variables, $\beta$ is the coefficient vector of order $(p+1) \times 1$ with intercept and $z_i$ is a random variable following the gamma distribution such that $z_i \sim \Gamma(\delta, \delta)$, $i = 1, 2, ..., n$.

The density function of the dependent variable $y_i$ is given by

$$\Pr(y = y_i | x_i) = \frac{\Gamma(\theta^{-1} + y_i)}{\Gamma(\theta^{-1})\Gamma(1 + y_i)} \left(\frac{\theta^{-1}}{\theta^{-1} + \mu_i}\right)^{\theta^{-1}} \left(\frac{\mu_i}{\theta^{-1} + \mu_i}\right)^{y_i}$$

where the overdispersion parameter $\theta$ is given as $\theta = 1/\delta$. The conditional mean and variance of the distribution are given respectively as follows:

$$E(y_i | x_i) = \mu_i,$$

$$Cov(y_i | x_i) = \mu_i(1 + \theta \mu_i).$$

The estimation of the coefficient vector $\beta$ is usually obtained by maximizing the following log-likelihood function

$$L(\theta,\beta) = \sum_{i=1}^{n}\left\{\left[\sum_{j=0}^{y_i-1}\log(j+\theta^{-1})\right] - \log(y_i!) - (y_i+\theta^{-1})\log(1+\theta\mu_i) + y_i\log(\theta) + y_i\log(\mu_i)\right\} \quad (1.1)$$

since $\log\left(\dfrac{\Gamma(\theta^{-1}+y_i)}{\Gamma(\theta^{-1})}\right) = \sum_{j=0}^{y_i-1}\log(j+\theta^{-1})$. The estimation of the parameter $\beta$ is usually obtained by the method of maximum likelihood estimation (MLE) which can be obtained by maximizing the equation (1.1) with respect to $\beta$, namely, solving the following equation

$$S(\beta) = \frac{\partial L(\theta,\beta)}{\partial \beta} = \sum_{i=1}^{n}\frac{(y_i-\mu_i)}{1+\theta\mu_i}x_i = 0. \quad (1.2)$$

Since the equation (1.2) is non-linear in $\beta$, one should use the following scoring method

$$\beta^{(r)} = \beta^{(r-1)} + I^{-1}\left(\beta^{(r-1)}\right)S\left(\beta^{(r-1)}\right) \quad (1.3)$$

where $S\left(\beta^{(r-1)}\right)$ is the first derivative of the log-likelihood function evaluated at $\beta^{(r-1)}$ and

$$I^{-1}\left(\beta^{(r-1)}\right) = E\left(\frac{\partial^2 L(X;\beta)}{\partial\beta\partial\beta'}\right) = X'W\left(\beta^{(r-1)}\right)X,$$

$W\left(\beta^{(r-1)}\right) = \text{diag}\left(\dfrac{\mu_i\left(\beta^{(r-1)}\right)}{1+\theta\mu_i\left(\beta^{(r-1)}\right)}\right)$ evaluated at $\beta^{(r-1)}$. In the final step of the algorithm, MLE of $\beta$ is obtained as follows:

$$\hat{\beta}_{MLE} = \left(X'\hat{W}X\right)^{-1}X'\hat{W}\hat{Z} \quad (1.4)$$

where $\hat{Z}$ is a vector with the ith element equals to $\log(\hat{\mu}_i) + \frac{y_i - \hat{\mu}_i}{\hat{\mu}_i}$, $\hat{W}$ and $\hat{\mu}_i$ are the values of $W(\beta^{(r-1)})$ and $\mu_i(\beta^{(r-1)})$ at the final step respectively, the hats show the iterative nature of the algorithm. This method is also known as the iteratively re-weighted least squares algorithm (IRLS).

However, when the matrix $X'\hat{W}X$ is ill-conditioned, i.e., the correlation between the explanatory variables are high, MLE becomes instable and its variance is inflated. This problem is called multicollinearity.

Although, applying shrinkage estimators is very popular in linear model to solve the multicollinearity problem (see [7], [9], [12], [10] etc.), count models have not been investigated in the presence of multicollinearity. Therefore, as an exception, [13] proposed to use the ridge regression [6] in negative binomial regression models. The negative binomial ridge regression estimator (RR) is obtained as follows:

$$\hat{\beta}_k = \left(X'\hat{W}X + kI\right)^{-1} X'\hat{W}\hat{Z}, k > 0 \tag{1.5}$$

where $I$ is the $(p+1) \times (p+1)$ identity matrix. The author proposed to use some existing ridge estimators to estimate the ridge parameter $k$.

Moreover, Liu estimator [11] is generalized to the negative binomial regression model in [14] and obtain the following negative binomial Liu estimator (LE)

$$\hat{\beta}_d = \left(X'\hat{W}X + I\right)^{-1} \left(X'\hat{W}X + dI\right)\hat{\beta}_{MLE}, 0 < d < 1. \tag{1.6}$$

Finally, motivated by the idea that combining the two estimators might inherit the advantages of both estimators, a two-parameter estimator which is a combination of RR and LE has been proposed in [8].

The purpose of this paper is to generalize Liu-type estimator [12] to the negative binomial regression and discuss some properties of the new estimator. The organization of the paper is as follows: In section 2, Liu-type negative binomial estimator (LT) is proposed, matrix mean squared error (MMSE) and mean squared error (MSE) properties are investigated and selection of the shrinkage parameters is discussed. In order to compare the performances of the estimators MLE, RR, LE and LT, a Monte Carlo simulation is designed and its results are discussed in section 3. A real data application is demonstrated to illustrate the benefits of LT in section 4. Finally, a brief summary and conclusion are provided.

## 2. New Estimator and MSE Properties

### 2.1. Construction of LT

Consider the linear regression model $Y = X\beta + \varepsilon$ where $X$ is an $n \times p$ data matrix, $\beta$ is the $p \times 1$ coefficient vector, $\varepsilon$ is the $n \times 1$ random error vector satisfying $\varepsilon_i \sim N(0, \sigma^2)$ and $Y$ is the $n \times 1$ dependent variable. When there is multicollinearity, the matrix $X'X$ becomes ill-conditioned and some of the eigenvalues of the matrix $X'X$ becomes close to zero and the condition number $\kappa = (\vartheta_{max} / \vartheta_{min})$ becomes very high such that $\vartheta_j, j = 1, 2, ..., p$ are the eigenvalues of $X'X$. Thus, the ordinary least square estimator (OLS), $\hat{\beta}_{OLS} = (X'X)^{-1} X'Y$ becomes instable. Therefore, [6] proposed ridge estimator $\hat{\beta}_{Ridge} = (X'X + kI)^{-1} X'Y$ which is obtained by augmenting $0 = k^{1/2}\beta + \varepsilon'$ to the original equation. However, large values of $k$ makes the distance between $k^{1/2}\beta$ and $0$ increase and to control the condition number, one should use large values of $k$ which imposes more bias to the ridge estimator. Therefore, Liu [12] proposed to augment $(-d / k^{1/2}) \hat{\beta}_{OLS} = k^{1/2}\beta + \varepsilon'$ to the original equation and obtain Liu-type estimator $\hat{\beta}_{k,d} = (X'X + kI)^{-1}(X'X - dI)\hat{\beta}$ where $k > 0, -\infty < d < \infty$ and $\hat{\beta}$ is any estimator. The results showed that $\hat{\beta}_{k,d}$ has a better performance than OLS and ridge estimator in the sense of MSE.

Therefore, it is a good idea to define a generalization of Liu-type estimator in negative binomial model to solve the problem of multicollinearity. In this study, a generalization of Liu-type estimator to the negative binomial regression model is proposed as follows

$$\hat{\beta}_{LT} = \left(X'\hat{W}X + kI\right)^{-1}\left(X'\hat{W}X - dI\right)\hat{\beta}_{MLE} \qquad (2.1)$$

$k > 0, -\infty < d < \infty$.

From the definition of LT, it is easy to see that LT is a general estimator including MLE, RR and LE as follows, $\lim_{d \to -k} \hat{\beta}_{LT} = \hat{\beta}_{MLE}$, $\lim_{d \to 0} \hat{\beta}_{LT} = \hat{\beta}_k$ which is the negative binomial ridge estimator (RR), $\lim_{k \to 1, d \to -1} \hat{\beta}_{LT} = \hat{\beta}_{LE}$ which is the negative binomial Liu estimator (LE).

In order to see the superiority of the estimator LT, MMSE containing all the relevant information regarding the estimators can be used as a comparison criterion. MMSE and MSE being the trace of MMSE of an estimator $\tilde{\beta}$ are respectively defined by

$$\begin{aligned} MMSE(\tilde{\beta}) &= E\left[(\tilde{\beta} - \beta)(\tilde{\beta} - \beta)'\right] \\ &= \mathrm{var}(\tilde{\beta}) + \mathrm{Bias}(\tilde{\beta})\mathrm{Bias}(\tilde{\beta})', \end{aligned} \qquad (2.2)$$

$$\begin{aligned} MSE(\tilde{\beta}) &= tr\left(MMSE(\tilde{\beta})\right) = E\left[(\tilde{\beta} - \beta)'(\tilde{\beta} - \beta)\right] \\ &= tr\left[\mathrm{var}(\tilde{\beta})\right] + \mathrm{Bias}(\tilde{\beta})'\mathrm{Bias}(\tilde{\beta}). \end{aligned} \qquad (2.3)$$

Thus MSE and MMSE of MLE are given by the following equations respectively

$$MSE(MLE) = tr(X'WX)^{-1} = \sum_{j=1}^{p+1} \frac{1}{\lambda_j}, \tag{2.4}$$

$$MMSE(MLE) = (X'WX)^{-1} \tag{2.5}$$

where $\lambda_j$ is the $j^{th}$ eigenvalue of the matrix $X'WX$.

There is a need to make a transformation in order to present the explicit form of the MMSE and MSE functions. Let $Q'X'WXQ = \Lambda = \text{diag}(\lambda_1, \lambda_2, ..., \lambda_{p+1})$ and $\alpha = Q'\beta$ where $\lambda_1 \geq \lambda_2 \geq ... \geq \lambda_{p+1} > 0$ and $Q$ is the matrix whose columns are the eigenvectors of the matrix $X'WX$.

Now, we obtain the bias and variance functions of the estimators to compute the MMSE and MSE functions easily.

The bias, variance, MMSE and MSE functions of LT, RR, and LE are obtained respectively as follows:

$$b_{LT} = \text{bias}(LT) = -(d+k)Q\Lambda_k^{-1}\alpha,$$

$$\text{var}(LT) = Q\Lambda_k^{-1}\Lambda_d^*\Lambda^{-1}\Lambda_d^*\Lambda_k^{-1}Q',$$

$$MMSE(LT) = Q\Lambda_k^{-1}\Lambda_d^*\Lambda^{-1}\Lambda_d^*\Lambda_k^{-1}Q' + (d+k)^2 Q\Lambda_k^{-1}\alpha\alpha'\Lambda_k^{-1}Q',$$

$$MSE(LT) = \sum_{j=1}^{p+1} \frac{(\lambda_j - d)^2}{\lambda_j(\lambda_j + k)^2} + \frac{(d+k)^2 \alpha_j^2}{(\lambda_j + k)^2}, \tag{2.6}$$

$$b_{RR} = \text{bias}(RR) = -kQ\Lambda_k^{-1}\alpha,$$

$$\text{var}(RR) = Q\Lambda_k^{-1}\Lambda\Lambda_k^{-1}Q',$$

$$MMSE(RR) = Q\Lambda_k^{-1}\Lambda\Lambda_k^{-1}Q' + k^2 Q\Lambda_k^{-1}\alpha\alpha'\Lambda_k^{-1}Q',$$

$$MSE(RR) = \sum_{j=1}^{p+1} \frac{\lambda_j}{(\lambda_j + k)^2} + \frac{k^2 \alpha_j^2}{(\lambda_j + k)^2},$$

$$b_{LE} = \text{bias}(LE) = (d-1)Q\Lambda_1^{-1}\alpha,$$

$$\text{var}(LE) = Q\Lambda_1^{-1}\Lambda_d \Lambda^{-1} \Lambda_d \Lambda_1^{-1} Q',$$

$$MMSE(LE) = Q\Lambda_1^{-1}\Lambda_d \Lambda^{-1} \Lambda_d \Lambda_1^{-1} Q' + (d-1)^2 Q\Lambda_1^{-1}\alpha\alpha'\Lambda_1^{-1}Q'$$

$$MSE(LE) = \sum_{j=1}^{p+1} \frac{(\lambda_j + d)^2}{\lambda_j (\lambda_j + 1)^2} + \frac{(d-1)^2 \alpha_j^2}{(\lambda_j + 1)^2},$$

where $\Lambda_k = \Lambda + kI$, $\Lambda_d^* = \Lambda - dI$, $\Lambda_1 = \Lambda + I$ and $\Lambda_d = \Lambda + dI$.

After computing the MMSE and MSE functions, LT is compared to the other estimators in the sense of MMSE in the following theorems by using the following lemma:

**Lemma 2.1** [4]: Let $M$ be a positive definite (p.d.) matrix, $\alpha$ be a vector of nonzero constants and $c$ be a positive constant. Then $cM - \alpha\alpha' > 0$ if and only if $\alpha'M\alpha < c$.

### 2.2. Comparison of LT versus MLE

The following theorem presents the condition that LT is superior to MLE:

**Theorem 2.2:** Let $(d+k)(2\lambda_j + k - d) > 0$, $j = 1, 2, ..., p+1$ and $b_{LT} = \text{bias}(\hat{\beta}_{LT})$. Then

$$MMSE(MLE) - MMSE(LT) > 0 \text{ iff } b_{LT}' \left[ \Lambda^{-1} - \Lambda_k^{-1}\Lambda_d^* \Lambda^{-1} \Lambda_d^* \Lambda_k^{-1} \right]^{-1} b_{LT} < 1.$$

**Proof:** The difference between the MMSE functions of MLE and LT is obtained by

$$MMSE(MLE) - MMSE(LT) = Q\left(\Lambda^{-1} - \Lambda_k^{-1}\Lambda_d^*\Lambda^{-1}\Lambda_d^*\Lambda_k^{-1}\right)Q' - b_{LT}b_{LT}'$$

$$= Q\operatorname{diag}\left\{\frac{1}{\lambda_j} - \frac{(\lambda_j - d)^2}{\lambda_j(\lambda_j + k)^2}\right\}_{j=1}^{p+1} Q' - b_{LT}b_{LT}'. \quad (2.7)$$

The matrix $\Lambda^{-1} - \Lambda_k^{-1}\Lambda_d^*\Lambda^{-1}\Lambda_d^*\Lambda_k^{-1}$ is p.d. if $(\lambda_j + k)^2 - (\lambda_j - d)^2 > 0$ which is equivalent to $\left[(\lambda_j + k) - (\lambda_j - d)\right]\left[(\lambda_j + k) + (\lambda_j - d)\right] > 0$. Simplifying the last inequality, one gets $(d + k)(2\lambda_j + k - d) > 0$. The proof is finished by Lemma 2.1. ∎

### 2.3. Comparison of LT versus RR

The following theorem gives the condition that LT is superior to RR:

**Theorem 2.3:**

Let $b = Q\Lambda_k^{-1}\alpha$ and $d < 2\min(\lambda_j)$. If $(d^2 + 2dk)b'\left[\Lambda_k^{-1}\Lambda\Lambda_k^{-1} - \Lambda_k^{-1}\Lambda_d^*\Lambda^{-1}\Lambda_d^*\Lambda_k^{-1}\right]^{-1}b < 1$, then $MMSE(RR) - MMSE(LT) > 0$.

**Proof:** The difference between the MMSE functions of RR and LT is obtained by

$$MMSE(RR) - MMSE(LT) = Q\left(\Lambda_k^{-1}\Lambda\Lambda_k^{-1} - \Lambda_k^{-1}\Lambda_d^*\Lambda^{-1}\Lambda_d^*\Lambda_k^{-1}\right)Q' + b_{RR}b_{RR}' - b_{LT}b_{LT}'$$

$$= Q\operatorname{diag}\left\{\frac{\lambda_j}{(\lambda_j + k)^2} - \frac{(\lambda_j - d)^2}{\lambda_j(\lambda_j + k)^2}\right\}_{j=1}^{p+1} Q' - (d^2 + 2dk)bb' \quad (2.8)$$

$$= Q\operatorname{diag}\left\{\frac{2d\lambda_j - d^2}{\lambda_j(\lambda_j + k)^2}\right\}_{j=1}^{p+1} Q' - (d^2 + 2dk)bb'.$$

Since $(d^2 + 2dk)bb'$ is nonnegative definite, it is enough to prove that $Q(\Lambda_k^{-1}\Lambda\Lambda_k^{-1} - \Lambda_k^{-1}\Lambda_d^*\Lambda^{-1}\Lambda_d^*\Lambda_k^{-1})Q' - (d^2 + 2dk)bb'$ is p.d. Now let $d < 2\min(\lambda_j)$, then using Lemma 2.1, the proof is finished. ∎

### 2.4. Comparison of LT versus LE

The following theorem presents the condition that LT is superior to LE:

**Theorem 2.4:** If $b_{LT}' \left[ \Lambda_1^{-1}\Lambda_d\Lambda^{-1}\Lambda_d\Lambda_1^{-1} - \Lambda_k^{-1}\Lambda_d^*\Lambda^{-1}\Lambda_d^*\Lambda_k^{-1} \right]^{-1} b_{LT} < 1$ and

$\lambda_j(k + 2d - 1) + d(k+1) > 0, 0 < d < 1$, then $MMSE(LE) - MMSE(LT) > 0$.

**Proof:** The difference between the MMSE functions of LE and LT is obtained by

$$MMSE(LE) - MMSE(LT) = Q(\Lambda_1^{-1}\Lambda_d\Lambda^{-1}\Lambda_d\Lambda_1^{-1} - \Lambda_k^{-1}\Lambda_d^*\Lambda^{-1}\Lambda_d^*\Lambda_k^{-1})Q' + b_{LE}b_{LE}' - b_{LT}b_{LT}'$$

$$= Q\mathrm{diag}\left\{ \frac{(\lambda_j - d)^2}{\lambda_j(\lambda_j + 1)^2} - \frac{(\lambda_j - d)^2}{\lambda_j(\lambda_j + k)^2} \right\}_{j=1}^{p+1} Q' + b_{LE}b_{LE}' - b_{LT}b_{LT}' \quad (2.9)$$

$$= Q\mathrm{diag}\left\{ \frac{\lambda_j(k + 2d - 1) + d(k+1)}{\lambda_j(\lambda_j + k)^2} \right\}_{j=1}^{p+1} Q' + b_{LE}b_{LE}' - b_{LT}b_{LT}'.$$

Similarly, since $b_{LE}b_{LE}'$ is nonnegative definite, it is enough to prove that $Q(\Lambda_1^{-1}\Lambda_d\Lambda^{-1}\Lambda_d\Lambda_1^{-1} - \Lambda_k^{-1}\Lambda_d^*\Lambda^{-1}\Lambda_d^*\Lambda_k^{-1})Q' - b_{LT}b_{LT}'$ is positive definite. Letting $\lambda_j(k + 2d - 1) + d(k+1) > 0, 0 < d < 1$, it is easy to see that Lemma 2.1 leads to the desired result. ∎

### 2.5. Estimating the parameters $k$ and $d$

The selection of shrinkage parameters in biased estimators has always been an important issue. There are various numbers of papers suggesting different types of estimation techniques of the ridge parameter, Liu parameter and so on. In this study, motivated by the work of [6], [9], [15], some methods to select the values of the parameters $k$ and $d$ are proposed.

Following [6], differentiating the equation (2.6) with respect to the parameter $k$, it is easy to obtain the following equation:

$$\frac{\partial MSE(\hat{\beta}_{LT})}{\partial k} = \sum_{j=1}^{p+1}\left(\frac{-2\lambda_j(\lambda_j+k)(\lambda_j-d)^2}{\lambda_j^2(\lambda_j+k)^4} + \frac{2(k+d)(\lambda_j+k)\hat{\alpha}_j^2 - 2(k+d)^2(\lambda_j+k)\hat{\alpha}_j^2}{(\lambda_j+k)^4}\right). \quad (2.10)$$

Simplifying the numerator of the above equation and solving for $k$, one can get the following individual estimators

$$\hat{k}_{LT}^j = \frac{\lambda_j - d(1+\lambda_j\hat{\alpha}_j^2)}{\lambda_j\hat{\alpha}_j^2}, \ j=1,2,...,p+1. \quad (2.11)$$

The condition $\lambda_j - d(1+\lambda_j\hat{\alpha}_j^2) > 0$ should hold to get a positive value of $\hat{k}_{LT}^j$. Thus, the following restriction

$$d < \frac{\lambda_j}{1+\lambda_j\hat{\alpha}_j^2} \quad (2.12)$$

should be satisfied.

Now, to estimate the parameter $k$, following [9] and using mean function, the following method is proposed:

$$\hat{k}_{AM} = \frac{1}{p+1}\sum_{j=1}^{p+1}\left(\frac{\lambda_j - d(1+\lambda_j\hat{\alpha}_j^2)}{\lambda_j\hat{\alpha}_j^2}\right) \quad (2.13)$$

which is the arithmetic mean of $\hat{k}_{LT}^j$.

Moreover, following [1], the maximum function is used to obtain the following estimator:

$$\hat{k}_{MAX} = \max\left(\frac{\lambda_j - d\left(1+\lambda_j \hat{\alpha}_j^2\right)}{\lambda_j \hat{\alpha}_j^2}\right). \tag{2.14}$$

After choosing the parameter $d$ using the equation (2.12), one can estimate the value of $k$ using one of the methods proposed. By plugging-in these estimates in LT, a better performance may be observed. In the following section, a Monte Carlo simulation is designed to compare the performance of the estimators for different scenarios.

To estimate the ridge parameter to be used in RR, Månsson [13] proposed different methods. In this study, $K5 = \max\left(\sqrt{\frac{\hat{\alpha}_j^2}{\hat{\sigma}_j^2}}\right)$ will be used in the simulation since the author reported that $K5$ is the best estimator in most of the situations investigated.

Moreover, some methods were proposed in [14] to choose the shrinkage parameter $d$ to be used in LE. However, $D5 = \max\left(0, \min\left(\frac{\hat{\alpha}_j^2}{1/\lambda_j + \hat{\alpha}_j^2}\right)\right)$ had the lowest MSE value in most of the situations. Therefore $D5$ is used to estimate $d$ in LE in the simulation study.

### 3. Monte Carlo Simulation Study

#### 3.1. Design of the Simulation

In the previous section, some theoretical comparisons are provided. In this section, an extensive Monte Carlo simulation study is designed to evaluate the performances of the estimators. Here is the description of the simulation.

Firstly, the observations of the explanatory variables are generated using the following equation

$$x_{ij} = \left(1-\rho^2\right)^{1/2} z_{ij} + \rho z_{ip} \qquad (3.1)$$

where $i = 1, 2, \ldots, n$, $j = 1, 2, \ldots p$, and $\rho^2$ represents the correlation between the explanatory variables and $z_{ij}$'s are independent random numbers obtained from the standard normal distribution.

The dependent variable of the NB regression model is generated using random numbers following the negative binomial distribution $NB\left(\mu_i, \mu_i + \theta \mu_i^2\right)$ where $\mu_i = \exp(x_i \beta), i = 1, 2, \ldots, n$. The slope parameters are decided such that $\sum_{j=}^{p} \beta_j^2 = 1$, which is a commonly used restriction in the field (see [9]).

In the design of simulation, three different values of $\rho$ corresponding to $0.90$, $0.95$, $0.99$ are considered. The value of $\theta$ is taken to be $1.0$ and $2.0$ due to [13]. Moreover, the following small, moderate and large sample size values are considered: $50$, $100$ and $200$. The numbers of explanatory variables are taken to be 4 and 6.

The simulation is repeated 2000 times, convergence tolerance is taken to be $0.00000001$ and the estimated MSE values of the estimators are computed as follows:

$$MSE\left(\tilde{\beta}\right) = \frac{\sum_{r=1}^{2000} \left(\tilde{\beta}_r - \beta\right)' \left(\tilde{\beta}_r - \beta\right)}{2000}, \qquad (3.2)$$

where $\tilde{\beta}_r$ is an estimator of $\beta$ at the rth replication.

### 3.2. Results of the Simulation

The estimated MSE values obtained from the Monte Carlo simulation are presented in Tables 1-2. It is observed from tables that the factors affecting the performance of the estimators are the value of $\theta$, the sample size $n$, the number of explanatory variables $p$ and the degree of correlation between the explanatory variables $\rho$.

According to the tables, increasing the value of $\theta$ makes an increase in the estimated MSE values. As the degree of correlation increases, MSE of MLE is inflated and MSE of RR is affected negatively. LT with $k_{AM}$ and $k_{MAX}$ show better performance than MLE and RR since an increase in the degree of correlation affects LT with $k_{MAX}$ slightly, i.e., LT with $k_{MAX}$ is the most stable estimator in the study. LE has also better performance than MLE and RR, however, LT with $k_{AM}$ and $k_{MAX}$ has the best performance in most of the situations considered.

Moreover, increasing the number of explanatory variables also affects the estimators negatively, i.e., their estimated MSE increases. Although high correlation makes an increase in the MSE of LT with $k_{MAX}$ when $p=4$, it becomes robust to the correlation when $p=6$. According to the results of the simulation, LT with $k_{MAX}$ has the best performance among the estimators.

Table 1. Estimated MSEs of the estimator when $p = 4$

| $\theta$ | 1.0 | | | 2.0 | | |
|---|---|---|---|---|---|---|
| $n$ | 50 | 100 | 200 | 50 | 100 | 200 |
| $\rho = 0.90$ | | | | | | |
| LT(kAM) | 0.2483 | 0.2095 | 0.1816 | 0.3782 | 0.2714 | 0.2446 |
| LT(kMAX) | 0.2632 | 0.2503 | 0.2444 | 0.3409 | 0.3242 | 0.3023 |
| RR | 1.3117 | 0.4823 | 0.2606 | 2.2441 | 0.8058 | 0.4131 |
| LE | 0.5761 | 0.3537 | 0.2164 | 0.7115 | 0.4952 | 0.3125 |
| MLE | 1.3640 | 0.4901 | 0.2629 | 2.4722 | 0.8298 | 0.4168 |
| $\rho = 0.95$ | | | | | | |
| LT(kAM) | 0.3517 | 0.2468 | 0.1925 | 0.5141 | 0.3377 | 0.2435 |
| LT(kMAX) | 0.3025 | 0.2956 | 0.2581 | 0.3892 | 0.3611 | 0.3178 |
| RR | 2.1664 | 1.2658 | 0.5033 | 3.3184 | 1.8066 | 0.8421 |
| LE | 0.7008 | 0.5668 | 0.3534 | 0.7977 | 0.6843 | 0.4894 |
| MLE | 2.2329 | 1.6571 | 0.5176 | 3.7717 | 2.8076 | 0.8691 |
| $\rho = 0.99$ | | | | | | |
| LT(kAM) | 0.7346 | 0.4317 | 0.2866 | 1.0059 | 0.7081 | 0.4235 |
| LT(kMAX) | 0.3556 | 0.3018 | 0.2635 | 0.3948 | 0.3743 | 0.3335 |
| RR | 4.9866 | 3.2202 | 2.5914 | 5.5368 | 3.5991 | 3.3719 |
| LE | 0.7128 | 0.7048 | 0.6975 | 0.8280 | 0.7363 | 0.7144 |
| MLE | 10.0215 | 6.1448 | 2.8633 | 18.1536 | 10.0081 | 4.7138 |

Table 2. Estimated MSEs of the estimator when $p = 6$

| $\alpha$ | 1.0 | | | 2.0 | | |
|---|---|---|---|---|---|---|
| $n$ | 50 | 100 | 200 | 50 | 100 | 200 |
| $\rho = 0.90$ | | | | | | |
| LT(kAM) | 0.3374 | 0.2339 | 0.1961 | 0.4678 | 0.2971 | 0.2497 |
| LT(kMAX) | 0.4309 | 0.3663 | 0.3056 | 0.4800 | 0.4148 | 0.3522 |
| RR | 2.1804 | 0.9271 | 0.4479 | 3.3428 | 1.4560 | 0.7134 |
| LE | 0.9627 | 0.5997 | 0.3586 | 1.2341 | 0.7702 | 0.5113 |
| MLE | 2.4655 | 0.9289 | 0.4501 | 4.3043 | 1.4674 | 0.7199 |
| $\rho = 0.95$ | | | | | | |
| LT(kAM) | 0.3933 | 0.2514 | 0.2020 | 0.5808 | 0.3451 | 0.2470 |
| LT(kMAX) | 0.3792 | 0.3271 | 0.2730 | 0.4477 | 0.3953 | 0.3411 |
| RR | 3.8504 | 1.8333 | 0.8282 | 5.2209 | 3.0453 | 1.5874 |
| LE | 1.0414 | 0.8734 | 0.5688 | 1.2917 | 1.1277 | 0.8297 |
| MLE | 5.1537 | 1.9955 | 0.8319 | 8.9456 | 3.5382 | 1.6650 |
| $\rho = 0.99$ | | | | | | |
| LT(kAM) | 0.7928 | 0.4866 | 0.2976 | 1.5414 | 0.7228 | 0.4800 |
| LT(kMAX) | 0.3701 | 0.3170 | 0.2655 | 0.5113 | 0.3851 | 0.3402 |
| RR | 10.3993 | 8.9291 | 3.9404 | 12.3918 | 11.8867 | 6.7377 |
| LE | 1.0980 | 1.0196 | 1.0905 | 1.5306 | 1.2021 | 1.1914 |
| MLE | 19.1744 | 10.6314 | 5.4518 | 36.6638 | 17.7556 | 8.4051 |

## 4. Real Data Applications

### 4.1. Sweden Traffic Data

In this subsection, we illustrate the benefits of new estimator LT using a real dataset. The dataset is taken from the official website of the Department of Transport Analysis in Sweden (www.trafa.se). A similar dataset is used in [14]. The dependent variable is the number of pedestrian killed and the explanatory variables are the number of kilometers driven by cars $X_1$ and trucks $X_2$. In this application, we try to investigate the effect of changing the usage of cars and trucks on the number of pedestrian killed. There are 21 different counties in Sweden and the data are pooled during the year 2013 for different counties. The condition number being the square root of the ratio of the maximum eigenvalue and the minimum eigenvalue of the data matrix is approximately 210.9146 showing that there is a moderate multicollinearity. The negative binomial regression model with intercept is estimated using IRLS algorithm for different estimators considered in this study. The results are reported in Table 3.

According to Table 3, the effect of increasing $X_1$ has a negative impact on the number of pedestrian killed which is not expected. It is known that the signs of coefficients may be wrong when there is multicollinearity. Moreover, the effect of increasing $X_1$ is low while the effect of increasing $X_2$ is high.

If we use biased estimators, the effect of increasing $X_1$ becomes positive which is expected and the effect of increasing is lower when compared to MLE. When we compare the standard errors of estimators, it is observed that LT with $k_{AM}$ and $k_{MAX}$ have lower standard errors than other estimators which makes them more stable. Thus, the estimator LT should be preferred since it has a lower standard errors compared to other estimators and meaningful coefficients compared to MLE.

Moreover, LT with $k_{AM}$ and $k_{MAX}$ have less MSE values than the other estimators. We also plot the MSE values of the estimators LT and RR for changing values of $k$ and LE for changing values of $d$ such that $0 < k < 1, 0 < d < 1$. We estimate the parameter $d$ using (2.12) for LT. According to Figure 1, we observe that when $0 < d < 0.16$ MSE of LE is smaller than MSE of LT. Otherwise LT has the least MSE value.

Table 3. Coefficients, standard errors and MSE values of estimators for Sweden traffic data.

|  | Coefficients | | | | |
|---|---|---|---|---|---|
|  | LT(kAM) | LT(kMAX) | RR | LE | MLE |
| $\beta_0$ | 2.3135 | 2.2615 | 2.3352 | 2.1943 | 2.3731 |
| $\beta_1$ | 0.6031 | 0.5395 | 0.4826 | 0.2423 | -0.6003 |
| $\beta_2$ | 0.9121 | 0.7123 | 1.1791 | 0.8522 | 2.5690 |
|  | Standard errors | | | | |
| $\beta_0$ | 0.3081 | 0.3010 | 0.3111 | 0.2920 | 0.3165 |
| $\beta_1$ | 0.6515 | 0.5017 | 0.9493 | 0.7479 | 3.3505 |
| $\beta_2$ | 0.6551 | 0.5055 | 0.9483 | 0.7451 | 3.3205 |
|  | MSE | | | | |
|  | 0.9493 | 0.6002 | 1.8975 | 1.2069 | 22.3520 |

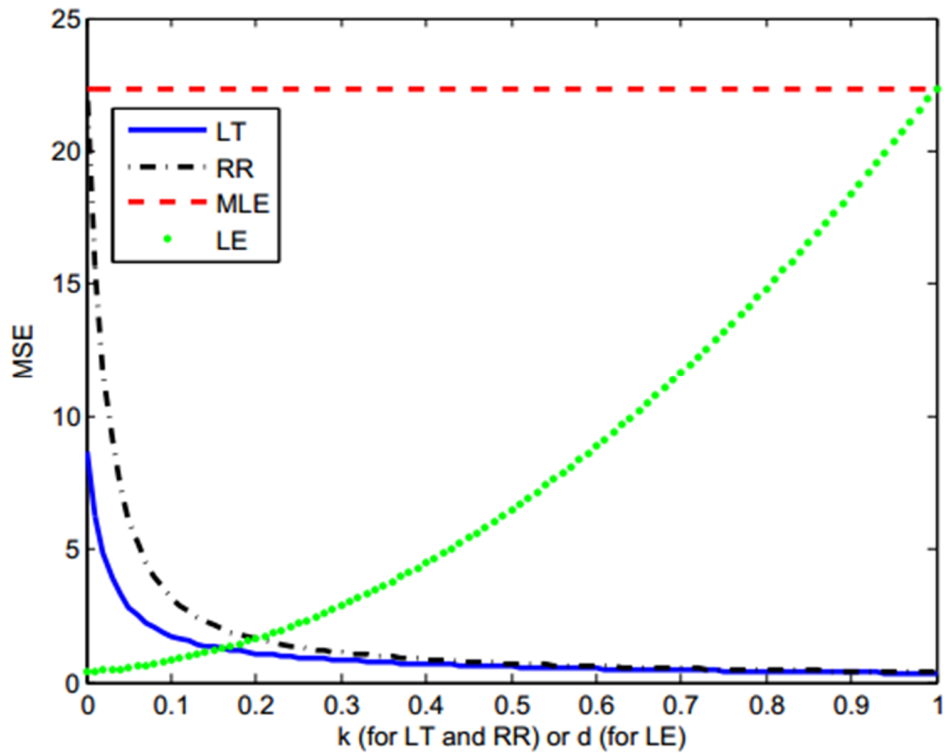

Figure 1. MSE plot of the estimators

Finally, we provide some information to justify the theorems given in Section 2. The estimated parameter values of LT are as follows: $k_{AM} = 0.2489$, $k_{MAX} = 0.4901$ and $d = 0.0192$. To justify Theorem 2.2, we consider the followings: $\min\left[(d+k)(2\lambda_j + k - d)\right] = 0.2881 > 0$ and $b_{LT}'\left[\Lambda^{-1} - \Lambda_k^{-1}\Lambda_d^*\Lambda^{-1}\Lambda_d^*\Lambda_k^{-1}\right]^{-1}b_{LT} = 2.8459e-05 < 1$ and the eigenvalues of the difference matrix $MMSE(MLE) - MMSE(LT)$ are $0.0071, 0.6927$ and $21.0520$ which are positive. Hence $MMSE(MLE) - MMSE(LT)$ is positive definite. Thus, Theorem 2.2 is satisfied.

Similarly, we compute the followings to justify Theorem 2.3, $2\min(\lambda_j) = 0.0947 > d = 0.0192$ $(d^2 + 2dk)b'\left[\Lambda_k^{-1}\Lambda\Lambda_k^{-1} - \Lambda_k^{-1}\Lambda_d^*\Lambda^{-1}\Lambda_d^*\Lambda_k^{-1}\right]^{-1}b = 7.2232e-08 < 1$ using $k_{MAX} = 0.4901$ for both RR and LT. The eigenvalues of the difference $MMSE(RR) - MMSE(LT)$ are $0.0002, 0.0203$ and $0.1058$ which are all positive, showing that the difference is positive definite. $(d^2 + 2dk)b'\left[\Lambda_k^{-1}\Lambda\Lambda_k^{-1} - \Lambda_k^{-1}\Lambda_d^*\Lambda^{-1}\Lambda_d^*\Lambda_k^{-1}\right]^{-1}b = 4.0861e-08 < 1$ for both RR and LT and again the difference is positive definite (the eigenvalues are $0.0003, 0.0346$ and $0.6454$). Thus Theorem 2.3 is satisfied.

Again, we consider the following computations to justify Theorem 2.4. We let $k = k_{AM} = 0.2489$ and $d$ is computed using (2.12) as $0.0192$ for both LE and LT. However, $\min\left[\lambda_j(k + 2d - 1) + d(k+1)\right]$ becomes negative and does not satisfy the pre-condition of Theorem 2.4. Thus, we try using $D5$ in both LE and LT to estimate the parameter $d$ which is computed as $0.1528$ and set $k = 1.1$. Now, $\min\left[\lambda_j(k + 2d - 1) + d(k+1)\right] = 0.0192 > 0$ which satisfies the pre-condition of Theorem 2.4. $b_{LT}'\left[\Lambda_1^{-1}\Lambda_d\Lambda^{-1}\Lambda_d\Lambda_1^{-1} - \Lambda_k^{-1}\Lambda_d^*\Lambda^{-1}\Lambda_d^*\Lambda_k^{-1}\right]^{-1}b_{LT} = 8.7066e-05 < 1$. The eigenvalues of the difference matrix $MMSE(LE) - MMSE(LT)$ are $0.0007, 0.1905$ and $0.5929$ which are all positive. Hence the difference matrix is positive definite.

Thus, we observe that Theorems given in Section 2 are satisfied.

### 4.2. Football Teams Data

In this subsection, another data set[1] regarding the football teams competing in the 2014-2015 Super League Season in Turkey is considered. A similar data set is also analyzed in [16] for the 2012-2013 season. According to [16], the data is appropriate for the Poisson regression model. However, we try to fit a negative binomial regression model because the variance (9.76) of the dependent variable is larger than the mean (7.33). Similar to their study, we have selected the number of won matches (NWM) as the dependent variable and the followings are the explanatory variables: the number of red cards (NRC), the number of substitutions (NS), the number of matches ending over 2.5 goals (NOG), the number of matches completed with goals (NCG), the ratio of the goals scores in number of matches [NGR1 = NGS/NM], and the ratio of goals scores in the sum of goals conceded and goal scores [NGR2 = NGS/(NGC + NGS)].

The eigenvalues of the data matrix are 900030, 99.2087, 58.6058, 30.2860, 0.4512 and 0.0646. The condition number is computed as $1.3928 \times 10^6$ which much larger than 1000 and shows that there is multicollinearity problem.

In Table 4, we present the coefficients and the standard errors of estimators. According to Table 4, it is easy to observe that the estimated theoretical MSE value of LT with $k_{AM}$ and $k_{MAX}$ are smaller than the others. Although, one can see that the variables NRC and NOG have negative impacts on NWM when RR, LE or MLE are used, this is not the case for the estimator LT. In other words, all the variables have positive but small effects on NWM when LT is used.

---

[1] Please see http://www.tff.org/default.aspx?pageID =164 and
    http://www.sahadan.com/takim_istatistikleri/Turkiye_Spor_Toto.

Moreover, the estimator LT has the least standard error values for this application which further shows the superiority of LT over the others.

Table 4. Coefficients, standard errors and MSE values of estimators for football teams data

|  | Coefficients | | | | |
|---|---|---|---|---|---|
|  | LT(k1) | LT(k2) | RR | LE | MLE |
| NRC | 0.0010 | 0.0008 | -0.0214 | -0.0171 | -0.0313 |
| NS | 0.0172 | 0.0160 | 0.0104 | 0.0068 | 0.0145 |
| NOG | 0.0019 | 0.0026 | -0.0696 | -0.0623 | -0.0801 |
| NCG | 0.0059 | 0.0045 | 0.0436 | 0.0779 | 0.0162 |
| NGR1 | 0.0005 | 0.0003 | 0.6473 | 0.2462 | 1.0650 |
| NG2 | 0.0003 | 0.0001 | 0.3805 | 0.1844 | 0.2735 |
|  | Standard errors | | | | |
| NRC | 0.0032 | 0.0008 | 0.1440 | 0.1378 | 0.1765 |
| NS | 0.0033 | 0.0029 | 0.0380 | 0.0358 | 0.0464 |
| NOG | 0.0039 | 0.0011 | 0.1223 | 0.1167 | 0.1527 |
| NCG | 0.0030 | 0.0010 | 0.1826 | 0.1572 | 0.2215 |
| NGR1 | 0.0004 | 0.0001 | 1.1622 | 0.3749 | 2.8818 |
| NG2 | 0.0002 | 0.0001 | 1.1713 | 0.3598 | 3.0555 |
|  | MSE | | | | |
|  | 0.0006 | 0.0094 | 2.7932 | 0.3286 | 17.7471 |

## 5. Conclusion

In this study, a new biased estimator which is a generalization of Liu-type estimator is proposed for the negative binomial regression models. We also review some existing estimators namely, negative binomial Liu estimator and negative binomial ridge estimator. We obtain some theoretical comparisons between the estimators using MMSE and obtain some conditions such that LT is superior to the others.

Moreover, we design a Monte Carlo simulation to understand the effects of the degree of correlation among the explanatory variables, the sample size and the number of explanatory variables. LT has a better performance than the others in the sense of MSE criterion in most of the cases considered in the simulation. Finally, we show that LT is a better choice and all the theoretical derivations are satisfied in real data applications and it is recommended to the researchers.